\documentclass[aps,prl,twocolumn,showpacs,preprintnumbers,amsmath,amssymb]{revtex4}

\usepackage{graphicx}
\usepackage{dcolumn}
\usepackage{bm}
\usepackage{subfigure}

\begin{document}

\title{Equivalent Dynamics from Disparate Synaptic Weights in a Prevalent Visual Circuit}

\author{Matthew S. Caudill}
\email{mcaudill@physics.wustl.edu}
\author{Sebastian F. Brandt}
 \author{Zohar Nussinov}
 \author{Ralf Wessel}
\affiliation{
Department of Physics, Campus Box 1105, Washington University in St. Louis, Missouri 63130-4899, USA
}

\date{\today}

\begin{abstract}
Neural feedback-triads consisting of two feedback loops with a non-reciprocal lateral connection from one loop to the other are ubiquitous in the brain. We show analytically that the dynamics of this network topology are determined by two algebraic combinations of its five synaptic weights. Thus different weight settings can generate equivalent network dynamics. Exploration of network activity  over the two-dimensional parameter space demonstrates the importance of the non-reciprocal lateral connection and reveals intricate behavior involving continuous transitions between qualitatively different activity states. In addition, we show that the response to periodic inputs is narrowly tuned about a resonant frequency determined by the two effective synaptic parameters.
\end{abstract}
                 
\maketitle
Neural microcircuits are well-defined networks of neurons with connectivity patterns exquisitely adapted for performing specific signal-processing tasks \cite{SheperdBook,Cabeza2006,GrillnerBook}. One prevalent connectivity pattern found among neural microcircuits consists of two feedback loops with a non-reciprocal lateral connection from one loop to the other (Fig.~1). We refer to this topology as a feedback-triad. In visual pathways for instance, the cortico-thalamic feedback system of mammals (Fig.~1a) \cite{ShermanBook2006,SillitoJonesFeedback} and the isthmotectal feedback system of birds and reptiles (Fig.~1b) \cite{GrubergMarin2006,WangKarten2004,WangKarten2006,Sereno1987,Powers1993} represent examples of  this network topology. 

In this Letter, we address the fundamental question of how synaptic weights influence the dynamics and signal-processing characteristics of the feedback-triad.  We show that the network's dynamics are not controlled by individual synaptic weights but rather by two algebraic combinations of them that crucially depend upon the non-reciprocal lateral connection for the generation of complex network activity. We further show that these two algebraic combinations determine the resonant frequency of the resonance profile of the network. The insights gained into how synaptic weights control a network's dynamics and signal-processing carry fundamental implications for our understanding of neural development, neuromodulation of network activity, and animal-to-animal variability of synaptic parameters.  

\begin{figure}
\begin{center}
\includegraphics[scale=.6]{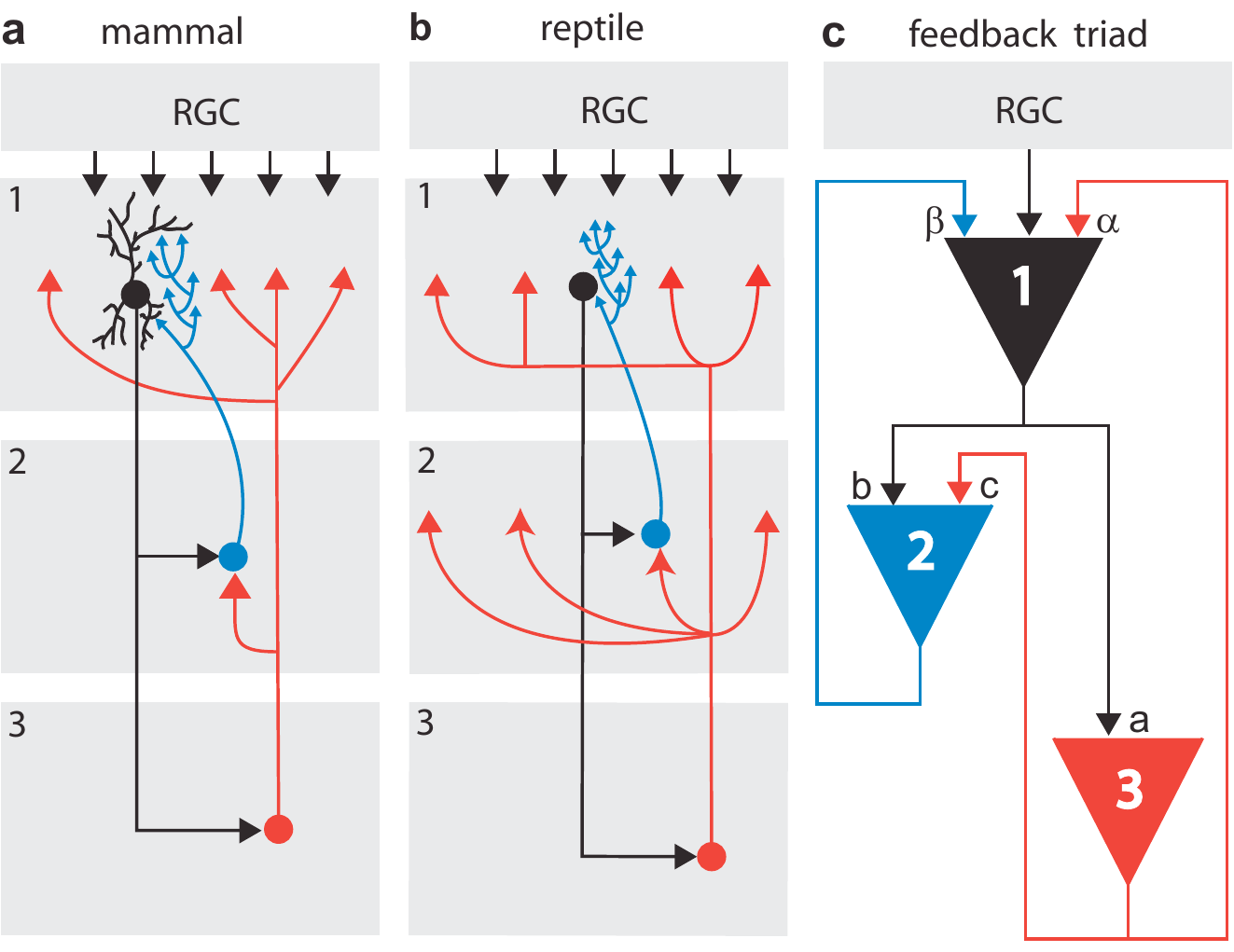}
\caption{\label{FIG. 1.} (Color) Neural feedback-triads. (a) In mammals, retinal ganglion cell (RGC) axons (black arrows) project to the thalamic lateral geniculate nucleus (1); this in turn projects to the thalamic reticular nucleus (2) and to the cortex (3). The latter two nuclei feed back to the lateral geniculate nucleus (1). (Modified after \cite{SillitoJonesFeedback}.) (b) In reptiles, RGC axons project to the optic tectum (1); this in turn projects to the nucleus isthmi pars parvocellularis (2) and to the nucleus isthmi pars magnocellularis (3). The latter two nuclei feed back to the optic tectum (1).  (Modified after \cite{Sereno1987}.) Note the non-reciprocal lateral connection from (3) to (2) in each circuit. (c) The feedback-triad consists of two feedback loops with a non-reciprocal lateral connection between them. }
\end{center}
\end{figure}

\begin{figure*}
\begin{centering}
\includegraphics[scale=.75]{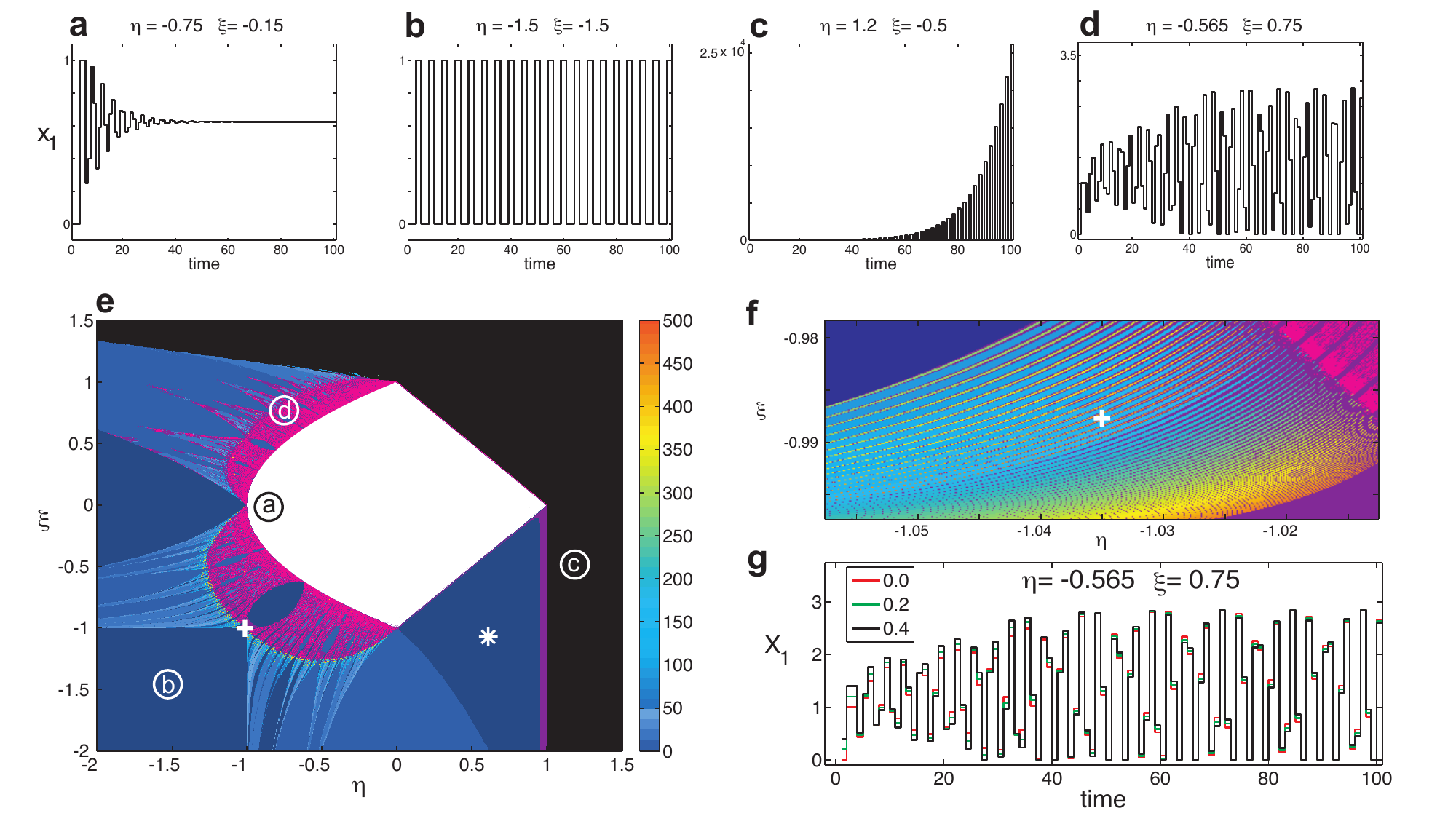}
\caption{\label{FIG. 2.} (Color) Dynamics of feedback-triads. Subfigures (a-d) show converging, oscillatory, diverging and aperiodic activity states that result from choosing different combinations of the two effective synaptic parameters $\eta$ and $\xi$. Subfigure (e) shows the color coded parameter space of activities. The white region corresponds to stable fixed points, the black region to unstable fixed points, the blue to orange colors represent oscillations with periods corresponding to the color and the cyan represents aperiodic states. Lastly, the purple region represents long period oscillations not represented on the color bar. (f) Zoom centered on the cross-hair indicating a region where small synaptic changes can result in different network activity. (g) Coexistence of multiple attractors (mulit-stability) at one combination of $\eta$ and $\xi$ accessed by different initial conditions. }  \end{centering}
\end{figure*}

The feedback-triad microcircuit (Fig.~1c) consists of three neurons with one input and five synaptic connections with identical delays $\tau$. For simplicity, we chose a discrete-time recurrent neural network with a piecewise-linear activation function \cite{Tsoi1997,Yi2004,Haschke2005} for the representation of the triad circuitry.  Specifically, the activity $x_i$ of neuron $i$ at time $t$ depends upon the activities $x_j$ of coupled neurons $j$ at the earlier time $t-\tau$ according to the following equations of motion  
 \begin{equation}
x_{i}(t)=\sigma \left( \sum_{j=1}^3w_{ij}x_{j}(t-\tau)+s_i \right).
\end{equation}
The matrix $w_{ij}$ measures the strength of connections between neurons $i$ and $j$ and the nonlinear transfer function $\sigma$ is chosen to be the max  function $\sigma(u)={\rm max}\{0,u\}  \quad u \in \Re$, guaranteeing nonnegative neuronal activities. Based on the circuit connectivity shown in Fig.~(1c), the synaptic weight matrix is given by 
\begin{equation}
 w= \left( \begin{array} {ccc} 0 & \beta & \alpha \\ b & 0 & c \\ a & 0 & 0
   \end{array} \right),
\end{equation}
where the feed-forward connections are denoted by latin letters and the recurrent connections by greek letters. Lastly, the input $\bm s$ is given by $\left(s_1,0,0\right)$ where $s_1$ is the input from the RGC axons.

We investigate the case in which the feed-forward weights $a$ and $b$ are positive because this assignment has been experimentally established in the vertebrate visual system \cite{Sherman2002}. In addition, for simplicity we assume that the weight of the lateral connection $c$ is also positive. Thus, the activities of neurons 2 and 3 will always be positive, and passage of the neuronal input through the nonlinear transfer function is unnecessary for these neurons. With this simplification, Eq.~(1) can be reduced from three equations to the following 3$\tau$-cycle difference map for the activity of the network, taken as the activity of neuron 1.
 \begin{equation}
 x_1(t) = \sigma \left[s_{1}+\eta x_1(t-2\tau)+\xi x_1(t-3\tau)\right].
 \end{equation}
The synaptic weights now appear only in the combinations
\begin{eqnarray}
\eta = \beta b +\alpha a, \quad \quad \xi=\beta a c,
\end{eqnarray} 	 
thereby reducing the complexity of the model from five synaptic parameters to two effective synaptic parameters.

In the case of instantaneous signal transfer (i.e. $\tau=0$) the network activity is stationary for all parameter values $\eta$ and $\xi$. However, for nonzero delay the dynamical behavior can be complex and is qualitatively independent of $\tau$ since Eq.~(3) can be rescaled in time. Thus we identify the delay with a unit time step and choose the network input $s_1=1$. By varying the effective synaptic parameters four types of dynamical behaviors emerge, namely, convergent (Fig.~2a), oscillatory (2b), divergent (2c), and aperiodic (2d). Color-coding these four activity states, a dynamical phase diagram representing network activity numerically in the $\eta$-$\xi$ phase plane has been obtained (Fig.~2e,f). This parameter space features qualitatively different activity-state regions with fascinating geometrical boundaries. One immediate result is that if the synapse $c$ is zero, the parameter $\xi$ vanishes leaving only converging and period-4 oscillations (Fig.~2e) as the biologically relevant activity states. Thus, the non-reciprocal connection between the feedback loops is crucial for generating complex network activity patterns. 

To understand the structure of the region of convergence, we first note that if the $\sigma$ operation is not performed in Eq.~(3) the size and shape of this dynamical region remains unchanged. Accordingly, the linearized version of Eq.~(3) written in standard form 
\begin{equation}
 x_1(t+3)-\eta x_1(t+1)-\xi x_1(t)-1=0,
 \end{equation}
should yield the boundaries of the convergent region. Assuming a solution of the form $x_1(t)=\lambda^t, $ Eq.~(5) yields three eigenvalues, one real and a complex-conjugate pair. The stability of the fixed points in this region is guaranteed when the moduli of these eigenvalues are less than one. By imposing this condition we determine the boundaries of the stable fixed-point region to be $\eta^2=\xi^2-1$, $\xi=\eta-1$, and $\xi=1-\eta$. These fixed-point boundary results are consistent with what is found for differential-delay-equation models studied previously \cite{Belair1996}. 

The structure of the limit-cycle regions in the parameter space cannot be obtained from the linearized form of Eq.~(1). Instead, they are a result of the nonlinearity imposed on the activities of the neurons by the transfer function. However, initial transients make it difficult to predict when the nonlinearity intervenes. To address this problem, we developed a set of constraint equations based on the steady-state limit-cycles obtained from our numerical simulation. Since the trajectory is periodic the activities of each neuron can be represented by a vector whose length $T$ is the period of the limit-cycle. The steady-state numerical results determine the positions of the zeros in each of the vectors. Elements from different vectors are related to each other by Eq.~(1), thus giving $3T$ constraint equations.  When solved simultaneously,  these equations yield polynomials in $\eta$ and $\xi$ that form the boundaries of the limit-cycle regions. For example, the period-2 region denoted by a star in Fig.~(2e), has six constraint equations which, when solved, yield the boundaries $\xi \leq \eta-1$ and $\eta \leq 1$. Neuromodulation of the synaptic weights would allow the network parameters to cross these boundaries, reconfiguring the circuit to produce different temporal patterns.

On applying this procedure to higher periods, we found that the structure of the striations in the parameter space are described by geometric series in the effective synaptic parameters. For example, in the upper-left quadrant of the parameter space, the upper and lower boundaries of each of the striations are respectively given by
\begin{subequations}
\label{eq:whole}
\begin{equation}
\sum_{i=1}^p{i\eta \xi^{i-1}}+\sum_{i=1}^{2p}{\xi^i}=0,
\label{subeq:1}
\end{equation}
and
\begin{equation}
1+\eta \left( \sum_{i=1}^p{i\eta \xi^{i-1}}+\sum_{i=1}^{2p}{\xi^i}\right)=0,
 \label{subeq:2}
 \end{equation}
 \end{subequations}
where $p$, the order of the polynomial, is related to the period of the region between an upper and a lower striation by the equation $p={(T-1)}/{3}$. 
The boundary separating the limit-cycles from the divergent states in the upper-left quadrant of the parameter space is not analytic. Even so, we found that it is described by the intersection of the striation boundaries. Specifically, the intersection of polynomials of order $p$ and $p+1$ determine a point along this upper left boundary of the parameter space.

In addition to stable (i.e. fixed-point) activity states and limit-cycles, the system also exhibits aperiodic states (Fig.~2d). We tested these states for chaotic dynamics by varying the initial conditions and checking whether the trajectories diverged from one another. The trajectories are not chaotic, but they do exhibit qualitatively different activity depending upon the initial conditions (Fig.~2g). This coexistence of multiple attractors is referred to as multistability and has been proposed as a mechanism for memory storage and pattern recognition \cite{Canavier1993,Wilson1972}.
\begin{figure}[h]
\begin{centering}
\includegraphics[scale=.45]{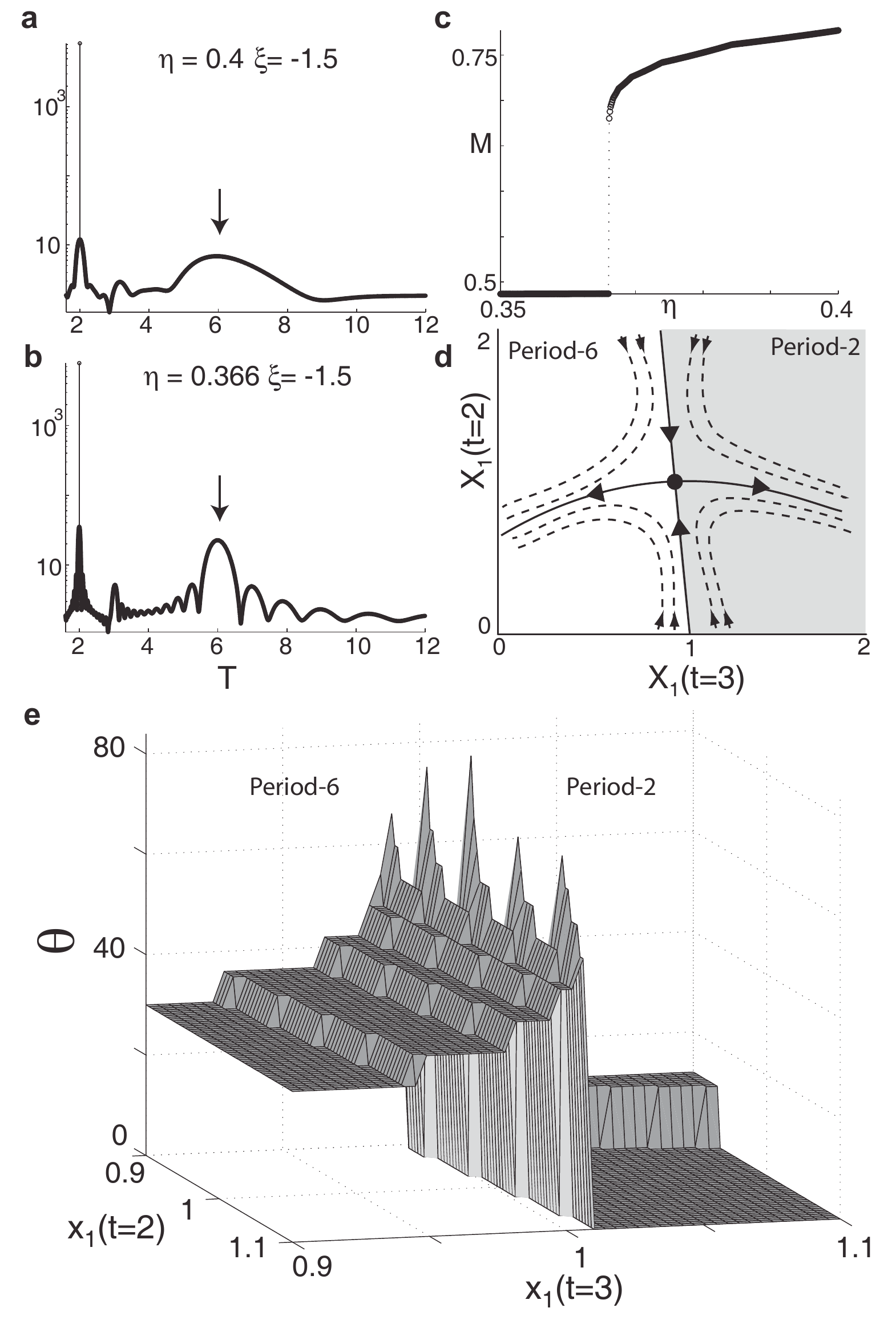}
\caption{\label{FIG. 3.}  Critical behavior near the boundary separating the period-2 and period-6 activity states. Plots (a,b) show the gathering of Fourier weights around period-6  as the path moves closer to the transition boundary. (c) Order parameter $M$ measured along this path shows a second-order phase transition. (d) Phase-space plot depicting activity responses for different $x_1$ initial conditions for a point on the boundary. (e) Transitory time $\theta$ measured near the critical value in the initial conditions.}  
\end{centering}
\end{figure}

The parameter space has many regions in which small changes in the effective synaptic weights can lead to qualitatively different network activity states. To understand the transition from one activity state to another, we employed an order parameter $M$ involving the Fourier transform of the network activity. This was motivated by the observation that the Fourier transform of the activity near a regional boundary contained  Fourier components consistent with periods corresponding to that region and the adjacent region (Fig.~3a,b).  The order parameter $M=1-{ \sum_{k' \neq k_{max}} I_{k'}}/{ \sum_{k}I_k}$ compares the power $I_k$ of each peak $k$ in the Fourier transform relative to the total power of the spectrum. As the boundary separating the period-2 (star in Fig.~2e) and period-6 activity states is crossed, a second-order phase transition occurs (Fig.~3c), indicating that points along the boundaries separating different network states will be sensitive to the initial network activity. To quantify this critical behavior, we determined numerically the network activity for various initial conditions (Fig.~3d) and measured the length of the transitory time $\theta$ before the network settles into steady-state activity (Fig.~3e). The results show that the boundaries are multistable, a feature that allows the circuit to show activity-dependent responses, such as are observed experimentally in the mammalian visual system \cite{SteriadeBook2006}. In addition, the boundaries of various regions exhibit a devil's staircase, (Fig.~3e), a self-similar fractal structure with scaling indexes universal to a large class of dynamical systems  \cite{Bak1986}.

An important signal-processing characteristic is the susceptibility of the feedback-triad to periodic inputs with specific driving frequencies $\omega$. In Fourier space, the linear response $\tilde r(\omega)$ of the triad is related to the stimulus $\tilde s(\omega)$ by $\tilde r(\omega)=\tilde \chi(\omega) \tilde s(\omega)$, where $\tilde \chi(\omega)$ is the AC susceptibility. This relationship holds for sufficiently weak driving input. For example, at the point $(\eta,\xi) = (-1.5,-1.5)$,  the {\em neural susceptibility} profile was found to follow a scaled Lorentzian of the form 
\begin{eqnarray}
\tilde \chi = b  \Big[ \frac{(\Gamma/2)}{\pi((\Gamma/2)^{2} + (\omega - \omega_{0})^{2})} \Big],
\label{chieq}
\end{eqnarray}
where the (narrow) Lorentzian width $\Gamma= 2.5 \times 10^{-4} $, the scale factor
$b=.72725$, and the Lorentzian {\em neural resonant frequency}  $\omega_{0} = 0.2$ (Fig.~4). Throughout
this work, we set the time step $\tau$ to unity. For a general time step $\tau$, the values of both $a$ and $\omega_{0}$ quoted above are in units of $\tau^{-1}$. \begin{figure}[h]
\begin{centering}
\includegraphics[scale=.4]{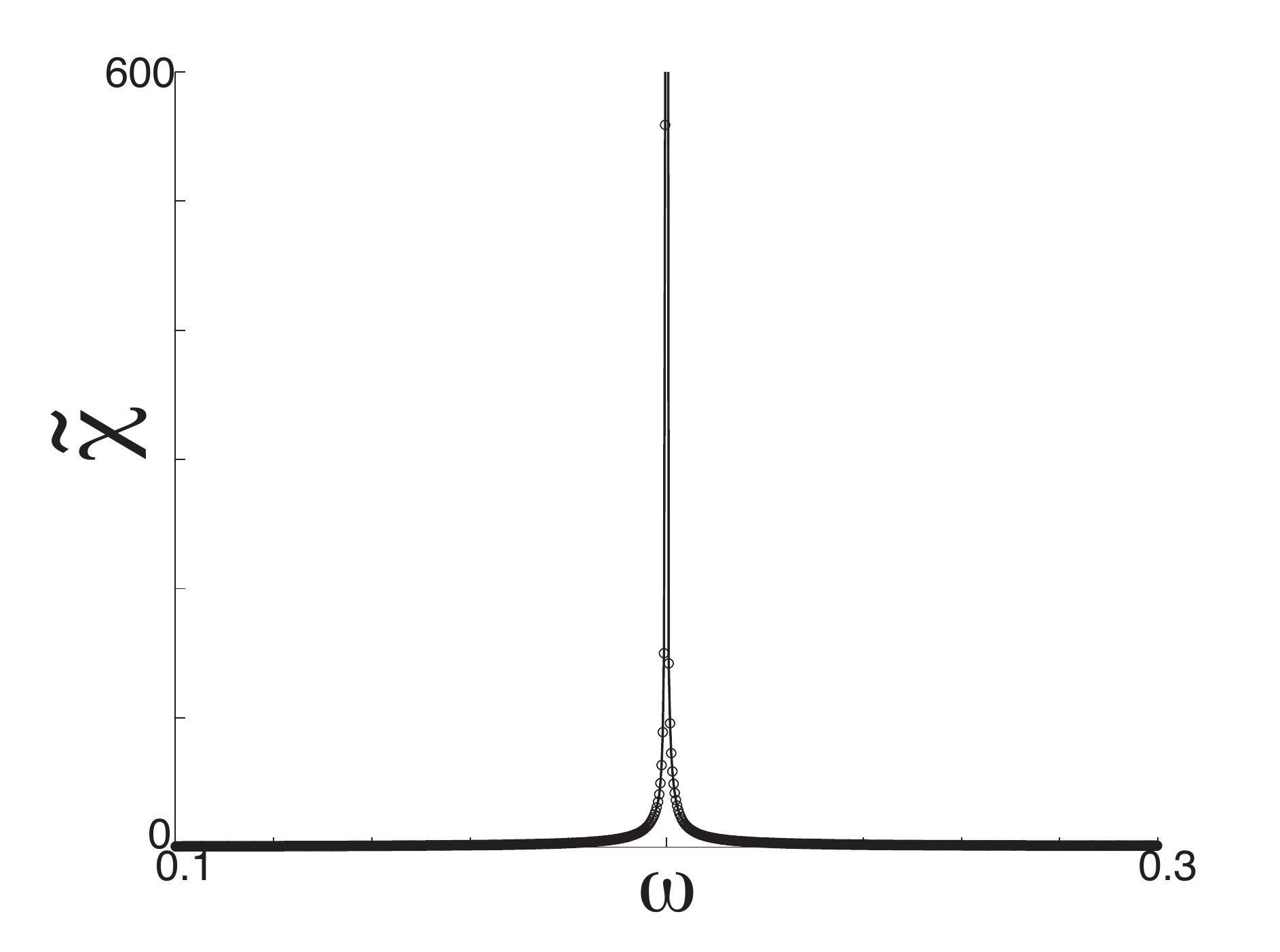}
\caption{\label{FIG. 4.} Susceptibility profile of the feedback-triad over frequencies given in units of $\tau^{-1}$. The susceptibility is sharply tuned about a resonant frequency determined by the two effective synaptic weights. See text and Eq.~(\ref{chieq}).}  
\end{centering}
\end{figure}
 This sharp tuning provides an effective bandpass filter that enables the extraction of a specific frequency from the input stimulus. Furthermore, since the resonant frequency depends on the values of the synaptic weights, neuromodulatory substances would allow the microcircuit to switch modes of preferred stimulus frequency. A similar analysis could be extended to larger networks comprised of feedback-triad microcircuits in ``series". For the case of $N$ feedback-triads the response would be related to the stimulus by $\tilde r(\omega)= \left( \prod_i^N \tilde \chi_i(\omega)\right) \tilde s(\omega)$ where $\chi_i(\omega)$ is the susceptibility of the $i^{th}$ feedback-triad.

Numerous neural and biochemical \cite{Alon2006,Papin2004} networks can be mapped onto the feedback-triad network topology we have investigated. Importantly, it has been shown that the dynamics and signal processing characteristics of this topology are determined by two combinations of the synaptic weights that depend upon the lateral connection for the generation of complex dynamics. The implications of this study are fourfold. First, the seeming incongruity in animal-to-animal variability of synaptic weights and equivalent dynamics is addressed because network activity is not determined by individual synapses but rather by certain combinations of the synaptic weights. Second, manipulation of individual synapses by biochemical agents must likewise be envisioned in terms of combinations of synaptic weights. Third, the existence of continuous transitions between different activity states in the two-dimensional parameter space highlights the flexibility of the dynamics in terms neuromodulation of the synaptic weights and activity-dependent responses. Last, the signal response is sharply tuned around a natural resonant frequency determined by the two effective synaptic weights, indicating that larger networks composed of feedback-triad microcircuits may be suited to bandpass filtering of neural stimuli.
\begin{acknowledgments}
We thank A. Carlsson and J. W. Clark for critical reading of the manuscript. This work was supported by NIH-EY 15678.
\end{acknowledgments}

\end{document}